**On the exceptional temperature stability of ferroelectric $Al_{1-x}Sc_xN$ thin films**

Md Redwanul Islam[1], Niklas Wolff[1], Mohamed Yassine[2], Georg Schönweger[4], Björn Christian[2], Hermann Kohlstedt[4], Oliver Ambacher[2,5], Fabian Lofink[6], Lorenz Kienle[1], Simon Fichtner[3,6]

[1] Synthesis and Real Structure, Institute for Material Science, Kiel University, Kaiserstr. 2, D-24143 Kiel, Germany

[2] Power Electronics, Institute for Sustainable Systems Engineering, Freiburg University, Emmy-Noether-str. 2, D-79110 Freiburg, Germany

[3] Microsystems and Technology Transfer, Institute for Material Science, Kiel University, Kaiserstr. 2, D-24143 Kiel, Germany

[4] Nanoelectronics, Institute of Electrical Engineering and Information Engineering, Kiel University, Kaiserstr. 2, D-24143 Kiel, Germany

[5] Fraunhofer Institute for Applied Solid State Physics (IAF), Tullastr. 72, D-79108 Freiburg, Germany

[6] Fraunhofer Institute for Silicon Technology (ISIT), Fraunhoferstr. 1, D-25524 Itzehoe, Germany

Through its dependence on low symmetry crystal phases, ferroelectricity is inherently a property tied to the lower temperature ranges of the phase diagram for a given material. This paper presents conclusive evidence that in the case of ferroelectric $Al_{1-x}Sc_xN$, *low temperature* has to be seen as a purely relative term, since its ferroelectric-to-paraelectric transition temperature is confirmed to surpass 1100 °C and thus the transition temperature of virtually any other thin film ferroelectric. We arrived at this conclusion through investigating the structural stability of 0.4 - 2 µm thick $Al_{0.73}Sc_{0.27}N$ films grown on Mo bottom electrodes via *in situ* high-temperature X-ray diffraction and permittivity measurements. Our studies reveal the wurtzite-type structure of $Al_{0.73}Sc_{0.27}N$ is conserved during the entire 1100 °C annealing cycle, apparent through a constant *c/a* lattice parameter ratio. *In situ* permittivity measurements performed up to 1000 °C strongly support this conclusion and include what could be the onset of a diverging permittivity only at the very upper end of the measurement interval. Our *in situ* measurements are well-supported by *ex situ* (scanning) transmission electron microscopy and polarization and capacity hysteresis measurements. These results confirm the structural stability on the sub-µm scale next to the stability of the inscribed polarization during the complete 1100 °C annealing treatment. Thus, $Al_{1-x}Sc_xN$ there is the first readily available thin film ferroelectric with a temperature stability that surpasses virtually all thermal budgets occurring in microtechnology, be it during fabrication or the lifetime of a device – even in harshest environments.

Ferroelectrics (FEs) are an important class of functional materials where the direction of spontaneous polarization can be altered by external electric fields in a non-volatile fashion. They enable numerous applications in microelectronics – from neuromorphic computing over classical memory to sensors and actuators[1,2]. Ferroelectricity arises in materials that undergo a transition from a high symmetry (non-polar) towards a low symmetry (polar) phase when cooling below a transition temperature (i.e the Curie temperature, $T_C$) [3]. Thus, ferroelectricity is inherently a phenomenon that can only appear in the lower temperature ranges of a phase diagram. In addition, the Landau-Devonshire theory associates the temperature dependence of the free energy of a FE material with its negative quadratic term[3]. Therefore, materials with lower $T_C$ tend to have lower coercive fields, while at the same time only material that feature a coercive field that is lower than their breakdown strength qualify as FEs. Consequently, most existing FEs have a $T_C$ < 600 °C[4–6]. This also applies to the technologically most relevant thin film FEs which belong either to the perovskite- or fluorite-type material class and, from the point of application, is further complicated by the fact that gradual depolarization sets in well before $T_C$. Moreover, temperature stability is one of the most important indicators for the overall stability of a material over its lifetime – a fact which is regularly employed in highly accelerated life tests[7–9].

This letter therefore aims to reveal the temperature stability of $Al_{1-x}Sc_xN$ – a nitride whose remarkable FE properties were recently discovered and which promises to advance the integration of FE functionality to microelectronics through excellent technological compatibility, very high spontaneous polarization as well as good resistance to depolarization[10]. While the rough classification of $Al_{1-x}Sc_xN$ as a high temperature piezoelectric is supported by previous studies, these were limited to *ex situ* measurements at either still moderate temperatures < 600 °C or low – and thus not fully FE – ScN contents[4,11,12]. Therefore, we extended the considered temperature range to 1100 °C, by performing *in situ*, *in vacuo* X-ray diffraction (XRD) and permittivity measurements to assess the possible occurrence of phase transitions in $Al_{1-x}Sc_xN$ . These *in situ* measurements are complemented by pre and post polarization-electric-field (PE) and capacitance-voltage (CV) measurements to elucidate the polarizability and polarization stability after annealing at high temperatures as well as (scanning) transmission electron microscopy (S)TEM to study structure changes on the sub-µm scale. Thereby, we arrive at the conclusion that the temperature stability of $Al_{1-x}Sc_xN$ surpasses that of virtually all other thin film FEs that are currently being considered for usage in microelectronics and is therefore particularly suited for high temperature applications such as power or harsh-environment electronics.

$Al_{1-x}Sc_xN$ thin films with thicknesses of 400 nm and 2 µm and Sc contents of 27% were grown in an Oerlikon MSQ 200 multi-source pulsed-DC sputter chamber using a processes previously described[13]. AlN(100nm)/Mo(100 nm) bottom- and Mo(100 nm) top-electrodes were deposited in an Oerlikon Clusterline 200 II sputtercluster. Details regarding the AlN process have also been previously published[14], while Mo was deposited at 300 °C with a power of 1.5 kW at a gas flow of 25 sccm Ar. In order to allow contact to the bottom electrode, the $Al_{1-x}Sc_xN$ films were etched in TMAH by employing the upper Mo layer as a hard mask. Subsequently, the upper Mo-layer was retracted to define 500 µm × 500 µm and 1 mm × 1 mm square capacitors by using standard lithography and $H_3PO_4$ as the etchant. For samples whose top-electrodes degraded under temperatures of 1100 °C, Cr electrodes (50x50 µm, 80 nm thick) were subsequently deposited at a deposition rate of 0.2 Å/s by a bell jar thermal evaporation unit at 150 A. (Oerlikon Univex 300, background Pressure 1.7e-5 mbar).

The *in situ* high temperature (HT)-XRD experiments were performed under vacuum (~$10^{-3}$ mbar) by symmetrical θ-2θ scans in temperature intervals of 100 °C up to 1100 °C (stating at room temperature) using a graphite-dome-capped heating stage (AntonPaar, DHS 1100 domed hot stage with maximum temperature deviation of ± 2 °C for less than 1mm thick samples) inside a Rigaku SmartLab diffractometer (9kW, Hypix detector). The temperature ramp was set to 100 °C/min with roughly 60 minutes dwell time at the selected temperatures.

The θ-2θ scans were performed with Cu(K)-α radiation (λ = 1.54059 Å) with a Ge(220)x2 monochromator inserted at the incident side. θ-2θ peak areas were calculated using the integration feature of Origin Peak Analyzer. The average full-width at half-maximum (FWHM) and intensity of the reflection were calculated by fitting a Pseudo-Voight profile inside XRDfit (Python based open-source tool for XRD peak fitting[15]). The surface microstructure of the specimens and chemical composition analysis (Sc content) by energy dispersive X-ray spectroscopy (EDS) were performed on Zeiss Gemini ultra55 Plus Field Emission Scanning Electron Microscope (FE-SEM) equipped with an SSD-EDS detector (Oxford Instruments). Cross-section TEM specimens of the Mo/AlSc$_{0.27}$N (400 nm) system were prepared using the focused ion-beam (FIB) method (FEI DualBeam Helios600 FIB-SEM). The structural TEM analyses were performed on JEOL JEM-2100 (thermionic source LaB$_6$, acceleration voltage 200 kV) and Tecnai F30 (field emission gun, 300 kV) microscopes.

For the *in situ* electrical characterization, temperature dependent dielectric measurements were performed under vacuum (~10$^{-6}$ mbar) using a precision LCR meter (Agilent E4980A). The sample was heated to approximately 1000 °C in 25 °C increments using a boron nitride heater, so that at each temperature after settling, the relative permittivity has been measured in a frequency range from 1 kHz to 2 MHz at 0 V bias and an AC voltage of 1 V. These measurements were also performed during the subsequent cooling. *Ex situ* characterization of the polarization and capacitance hysteresis was performed with the help of an aixACCT TF 2000 analyzer.

This study focuses on 400 nm and 2 µm thick Al$_{0.73}$Sc$_{0.27}$N films deposited on AlN/Mo bottom electrodes and capped with Mo top electrodes. These films were selected for three specific reasons: First, their ScN content is high enough to result in fully saturated polarization during FE switching. Second, Mo is well known for its good temperature stability[16] and the texture of Al$_{1-x}$Sc$_x$N on Mo is generally sufficient to result in good piezoelectric properties, especially in films thinner than 1 µm[13] (however, the choice of Mo as the bottom interface does limit the amount of Sc that can be incorporated without degrading the texture[13]). Finally, besides being generally unwanted from the point of film performance, the occurrence of misaligned grains (SI Figure S2) with in-plane *c*-axis has one particular advantage for this work: Their presence allows the direct extraction of the lattice parameter *a* from symmetric θ-2θ scans through the appearance of the 10$\bar{1}$0 reflection. Therefore, both *c* and *a* lattice parameter can be measured under identical constraints (in-plane clamping by the substrate).

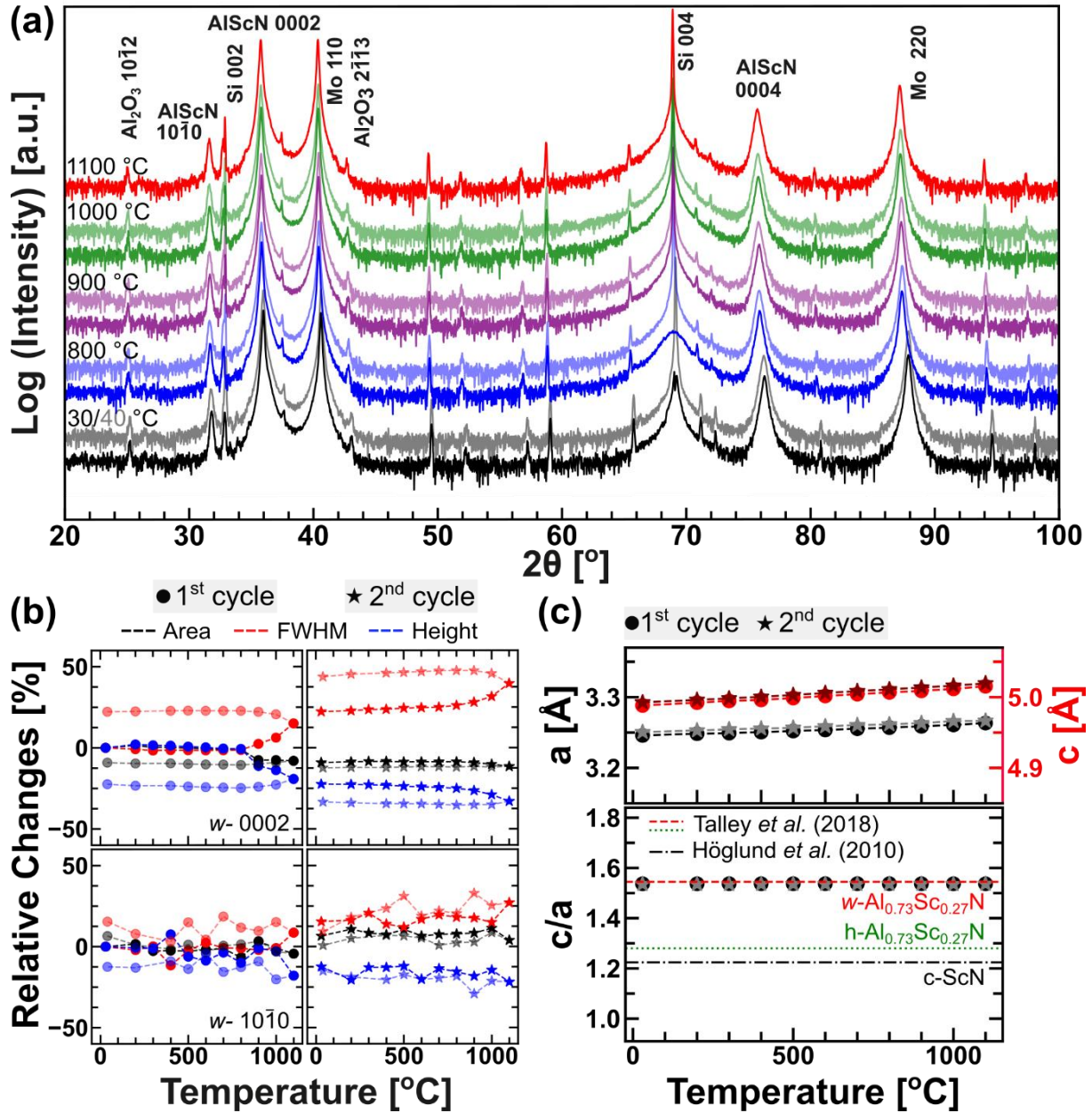

Figure 1: (a) *In situ* HT-XRD θ-2θ scans of Mo/Al$_{1-x}$Sc$_x$N (x = 27%), the corresponding cooling cycles are shown in semi-transparent colors (the complete cycle with all intermediate temperature steps are available as a supplement). (b) Relative changes of *w*-0002 and w-10$\bar{1}$0 reflection area, FWHM and height with temperature, relative to the scan at 30 °C. (c) Calculated a and c lattice parameters with their corresponding c/a ratio compared to literature (the theoretical data was not calculated for different temperatures and is only displayed as a line to guide the eye)[17].

The structural high-temperature stability of the Mo/Al$_{0.73}$Sc$_{0.27}$N thin film system was investigated by *in situ* HT-XRD with support of TEM inspection of the as-deposited and post-annealed specimen. The outcome of the *in situ* XRD experiment on the 2 μm thick Al$_{0.73}$Sc$_{0.27}$N film is plotted in figure 1 (a), showing the 1$^{st}$ heating and cooling cycle and figure 1 (b) the evolution of the *w*-0002 and *w*-10$\bar{1}$0 reflections in terms of intensity, integrated area and FWHM for the 1$^{st}$ and 2$^{nd}$ heating cycle (Figure S1). Qualitatively, no differences that could directly hint at large scale phase transitions can be identified from the diffraction patterns. Quantitatively, weak signs of degradation can be observed in the Al$_{0.73}$Sc$_{0.27}$N 0002-reflection area above 800 °C (about 10%), especially during the 1$^{st}$ heating cycle and to a significantly lower degree during the subsequent 2$^{nd}$ cycle on the same sample (figure 1 (c)). This observed degradation extends the overall result by Minghua *et al*. where an improvement of crystalline quality was reported at temperatures exceeding 800 °C[12] – albeit with only about half the Sc content and 6x lower dwell time compared to this work.

To rule out the possibility of any incipient diffuse FE-to-paraelectric phase transition, the $a$ and $c$ lattice parameters have been extracted from the $w$-10$\bar{1}$0 and $w$-0002 reflections of the 2 µm film (figure 1 (c)). While the thermal expansion of the material gives rise to the expected monotonous lattice expansion, the $c/a$ ratio remains virtually constant up to 1100 °C. Since both energetically closest competing phases (layered hexagonal (h), $P6_3/mmc$ as well as rocksalt-type (c), $Fm\bar{3}m$) have fundamentally smaller lattice parameter ratios, the existence of any large-scale phase transition can be excluded for the whole temperature range. Therefore, HT-XRD sets the lower limit for the Curie temperature of $Al_{0.73}Sc_{0.27}N$ at 1100 °C.

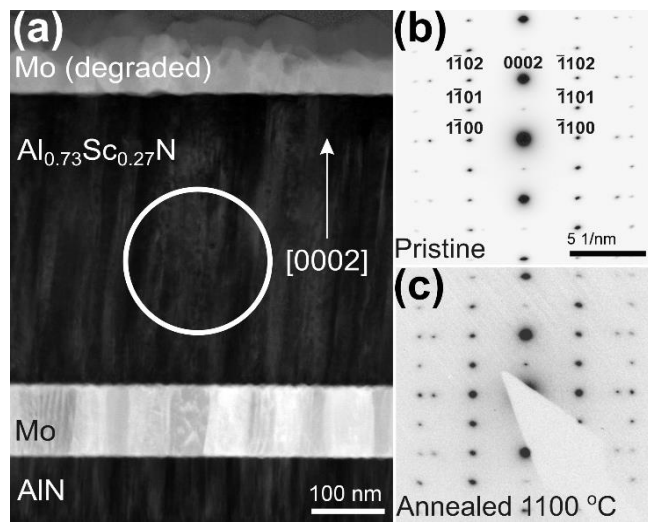

Figure 2: STEM HAADF image of the annealed specimen along with the comparison of the SAED patterns of the pristine and post-annealed specimen.

By employing (S)TEM, the overall integrity of the annealed $Al_{0.73}Sc_{0.27}N$ films was studied on the sub-µm scale. Cross-section specimens of as-deposited and *in situ* annealed 400 nm thick $Al_{0.73}Sc_{0.27}N$ thin films (*In situ* XRD with post-annealed SEM and Pole figures are available in SI Figure S3, S4) were compared by STEM using a high-angle annular darkfield detector and selected area electron diffraction (SAED) to identify apparent changes of the microstructure. The STEM micrograph of the annealed film is presented in figure 2 (a), next to the corresponding SAED pattern before and after annealing. Besides degradation of the top-Mo electrode, no qualitative changes of the microstructure with respect to the formation of secondary phases were observed with SAED comparing the annealed to the as-grown film (SI Figure S5). However, based on the large scale diameter of the analyzed circular area (diameter ~200 nm) the emergence of nanoscale secondary phases that would accompany e.g. the chemical segregation into Al and Sc rich phases[18] cannot be ruled out conclusively at this stage.

To further support the claim that $Al_{1-x}Sc_xN$ is a FE material with exceptional temperature stability, we conducted *in situ* permittivity measurement on the same 400 nm sample as discussed above. The permittivity values of this sample are shown in figure 3 (a) as a function of temperature, measured at different frequencies. From the curves in figure 3 (a), an unexpected diffuse dielectric anomaly can be seen. At temperatures between 350 and 900 °C, a broad and rounded peak is observed whose maximum decreases with an increase in frequency. Moreover, the position of the peak ($T_m$) also shows a frequency dispersion. In other polar materials, such peaks are either associated with relaxor behavior or charged defects, especially oxygen vacancies[19–21]. However, relaxor ferroelectrics also have low remnant polarization at RT[22,23] and often contain cations with different valences, which is not consistent with $Al_{1-x}Sc_xN$ as both Al and Sc cations have the same valency (3). Additionally, the material also has a low relative permittivity (10-30) and a large remnant polarization at RT[10]. In addition, relaxor FEs should exhibit a high symmetry phase at temperatures well beyond $T_m$, where the permittivity returns to a local minimum, and the Curie-Weiss law is valid[24]. This point should have been reached at 800 °C, however the HT-XRD measurements show no indication of such a large-scale loss of polarity above 800 °C. A defect-induced dielectric anomaly therefore remains as the only plausible explanation. This hypothesis is also supported by the cooling part of our curves, as the peaks are drastically reduced and disappeared at frequencies above 10 kHz, serving as indication of the suppression of defects through annealing. A similar diffuse dielectric anomaly has been observed in several perovskite-type oxide FEs at comparable temperatures and was associated to extrinsic phenomena such as the excitation of oxygen vacancies[20]. Nitrogen vacancies could lead to similar phenomena in $Al_{1-x}Sc_xN$ and the identification of the exact nature of the underlying defect will be the focus of future studies. Around 900 °C the curves for different frequencies largely follow the same path again and show a monotonous increase over all frequencies. This indeed might be the onset of divergence that can be associated with the true $T_C$. Thus, the *in situ* dielectric characterization of $Al_{0.73}Sc_{0.27}N$ fully in line with our microstructure analysis by HT-XRD and (S)TEM indicating structural stability of the material.

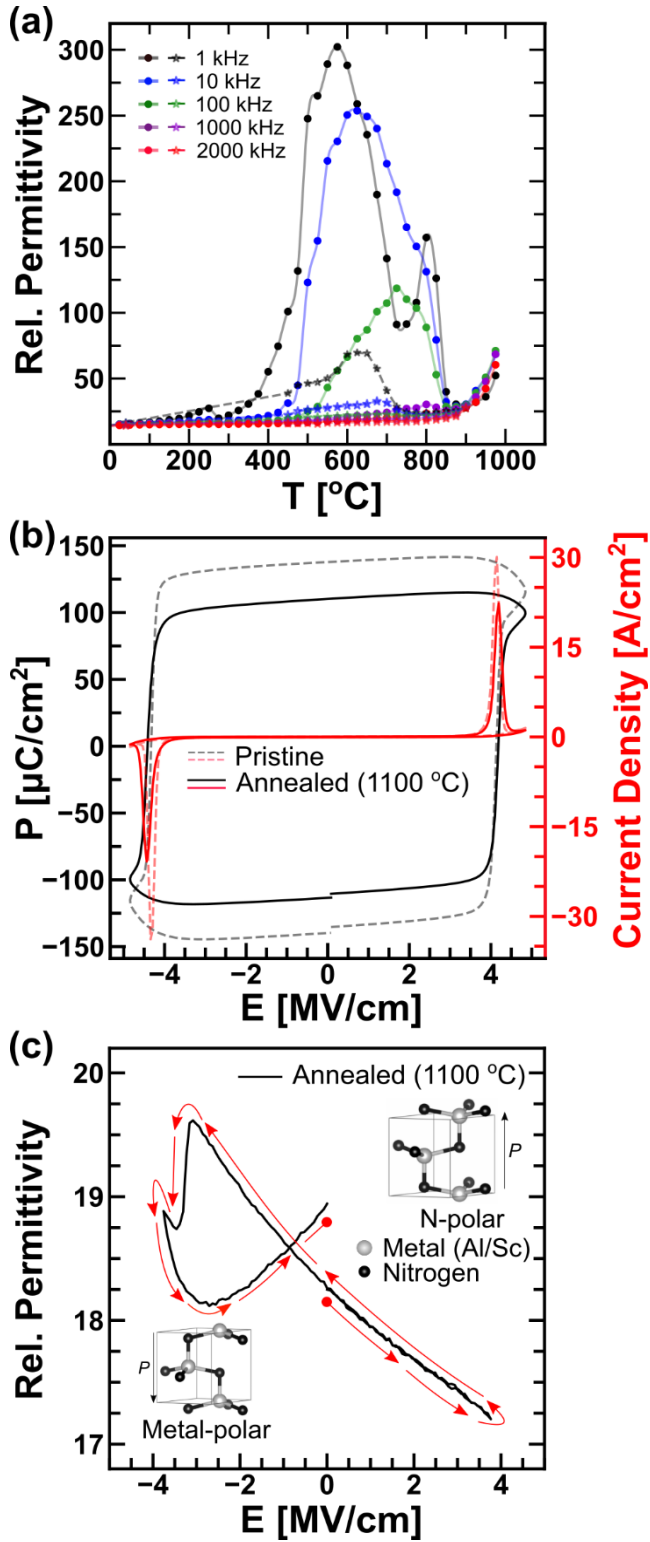

Figure 3: (a) *In situ* relative permittivity measurements at temperatures up to 1000 °C. The solid-colored lines are the interpolation of the data points at corresponding frequencies. Permittivity during cooling is marked in (*) with connecting dashed lines. (b) Polarization vs. electric field of as-deposited and post-annealed specimen (c) Relative permittivity of 1100 °C annealed $Al_{0.73}Sc_{0.27}N$ over bias (CV-measurement, the arrows indicate the applied bias cycle).

While the above experiments confirm that $T_C$ of $Al_{1-x}Sc_xN$ is well outside the range of what is typically required in microelectronics (even for extreme application scenarios), $T_c$ alone does not define the maximum use temperature of a FE. Degradation mechanisms such as delamination, the onset of segregation/decomposition, microcracking, hillock formation and oxidation due to insufficient inert ambient conditions can set in already at significantly lower temperatures. Therefore, the FE properties before and subsequent to the annealing experiment were investigated on the 400 nm AlN/Mo/AlS$_{0.73}$Sc$_{0.27}$N sample. Due to the degraded top-Mo layer (figure 1 (c)), a new set of Cr top-electrodes were deposited for electrical characterization. The resulting PE loops are compared in figure 3 (a). Clearly, the coercive field remains virtually constant at 4.1 MV/cm and also the qualitative box-like shape of the hysteresis is conserved – both suggesting that the local energy landscape is not significantly altered by the annealing cycle. The remanent polarization (P$_r$, uncorrected with respect to leakage) on the other hand degraded by around 20 % from approximately 138 µC/cm² to 110 µC/cm². The observed reduction of integrated area under the 0002 reflection and the conserved PE hysteresis during the heating cycle suggest, that the reduced polarization is coupled to the reduction of FE volume, in contrast to any structural changes within the material. Even the reduced polarization subsequent to the annealing step would however still be more than sufficient for e.g., application in a FE field effect transistor or random access memory[25].

Since gradual local depolarization might set in through an order-disorder transition or the presence of a depolarization field, the polarization retention during the annealing experiment was studied with the help of CV measurements (figure 3 (c)). Since this measurement was conducted without any intermediate repolarization, it reveals the overall polarization state of the Al$_{0.73}$Sc$_{0.27}$N film after annealing. Figure 3 (c) therefore clearly shows that the sample fully remained in its initial N-polar state – i.e., that no depolarization took place even at 1100 °C.

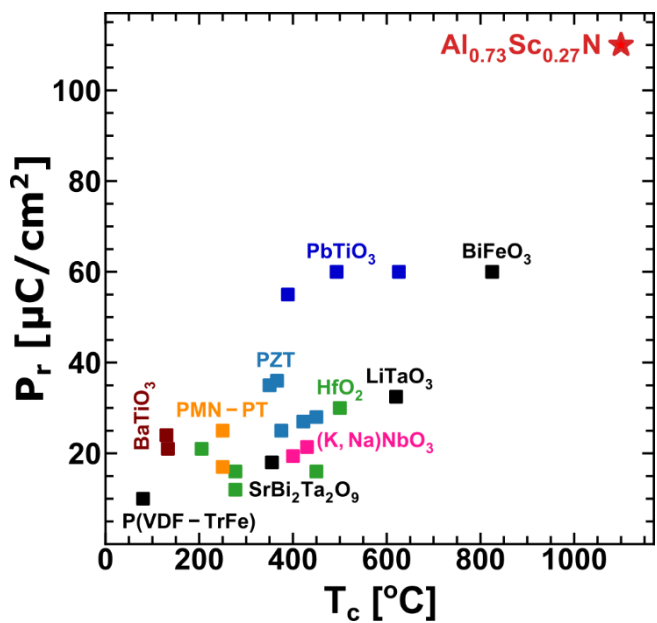

Figure 4: Comparison of the spontaneous polarization and lower limit of T$_C$ of Al$_{0.73}$Sc$_{0.27}$N with representatives of the more commonly considered thin film FEs: BaTiO$_3$[26,27], PZT[28,29], BiFeO$_3$[30,31], PbTiO$_3$[32], (K,Na)NbO$_3$[33], HfO2 based[5,34–36], PMN-PT: Pb (Mg$_{0.33}$Nb$_{0.67}$)O$_3$-PbTiO$_3$[37,38], LiTaO$_3$[39] and P(VDF-TrFe)[40] (Materials with multiple sources for Pr and Tc are marked in color).

To conclude, we could obtain comprehensive evidence that Al$_{1-x}$Sc$_x$N is a thin film ferroelectric with exceptional temperature stability compared to alternative materials (figure 4). *In situ* structural and dielectric analysis showed no sign of a ferroelectric to paraelectric phase transition up to 1100 °C. *Ex situ* (S)TEM analysis confirmed that structural changes on the sub-µm scale are moderate at most, as did the PE and CV hysteresis curves confirming the conservation of the inscribed polarization state during the entire annealing procedure. Together with its well-known compatibility to major technology platforms, this makes Al$_{1-x}$Sc$_x$N currently the most suitable FE for the realization of applications with challenging temperature budgets, be it during fabrication or lifetime, e.g., in the context of harsh environment memory.

**SUPPLEMENTARY INFORMATION**

The supplementary information provides all measurement steps of the HT-XRD scans of $Al_{0.73}Sc_{0.27}N$ (2 and 0.4 µm) along with SEM images of the pre- and post-annealed sample surface. For 400 nm $Al_{0.73}Sc_{0.27}N$ sample, it also contains X-ray pole figures (pristine and annealed), and a bright field TEM image (pristine).

**DATA AVAILABILITY**

The data shown in this work is available from the authors upon reasonable request.

**ACKNOWLEDGEMENTS**

This work was supported by the project 'ForMikro-SALSA' (grant no. 16ES1053) from the Federal Ministry of Education and Research (BMBF) and the Deutsche Forschungsgemeinschaft (DFG) under the scheme of the collaborative research center (CRC)1261. This work was also partially supported by the German Science Foundation (DFG) project No. AM 105/40-1 as well as by the Gibs-Schüle-Stiftung and the Carl-Zeiss-Stiftung (project 'SCHARF').

Phys. Rev. Lett. **84**, 175 (2000).